# Characteristic time of crossing a long free energy barrier

Alexei V. Finkelstein

Institute of Protein Research, Russian Academy of Sciences, 142290, Pushchino, Moscow Region, Russia;
E-mail: afinkel@vega.protres.ru

This short paper presents a simple approximate analytical estimate of the characteristic time of crossing a high, long and arbitrary bumpy free energy barrier in a course of chemical, biochemical or physical reaction.

Many important reactions, such as polymerization or aggregation, include crossing of a long free energy barrier [1-4] (Fig. 1).

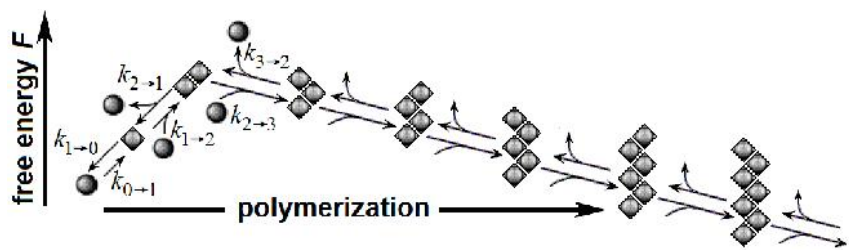

FIG. 1: Scheme of a polymerization reaction. Circles are free monomers in solution. Rhombs are monomers in a form appropriate for polymerization. This form is unstable for free monomers; very short polymers are also unstable; long polymers are stable. $k_{i \to i+1}$ is the transition rate constant for passing from state $i$ to the next state $i+1$ at the reaction pathway; $k_{i+1 \to i}$ is the rate constant for the reverse transition from $i+1$ to $i$ state.

Despite the abundance of such reactions (see, e.g., Fig. 2 as one more example), I, to my surprise, failed to find in literature a general formula to estimate their characteristic times.

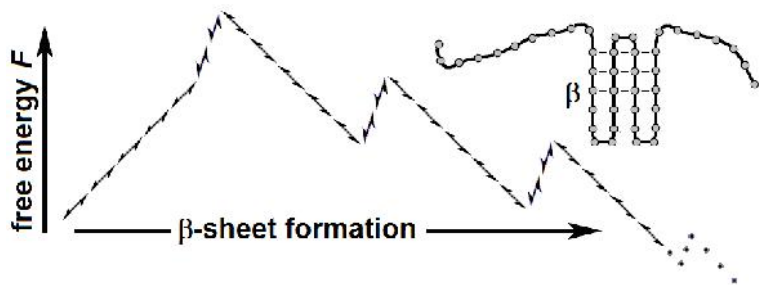

FIG. 2: Scheme of a free energy change along a pathway of β-sheet formation [5]. Each turn of the chain increases the chain's free energy, as well as the first extended chain region (separate β-strand), while each subsequent β-strand of the sheet decreases the free energy, so that a large β-sheet is stable.

It was not difficult to obtain an estimate of the characteristic time of crossing of a long, arbitrary bumpy but high free energy barrier (see equation (5) below), and this formula was included as equation (8.19) in [5] without a proof or analysis.

Here, I would like to prove and analyze this equation.

Consider the kinetics of a reversible process

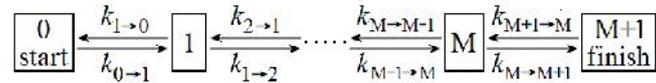

where $k_{i \to i+1}$ is the transition rate constant of passing from state $i$ to $i+1$, and $k_{i+1 \to i}$ of the reverse transition from $i+1$ to $i$ (we will assume that $F_0$, the free energy of the "start" state, is higher than that $F_{M+1}$, the free energy of the "finish" state).

The Moscow scientist Rakowski gave a general solution of the system of kinetic differential equations corresponding this process in 1907 [6]; but now we are interested in a simple estimate of the rate of such a specific process, in which the free energy of the intermediate states (1, …, M) is much higher than the free energy of the initial state 0 and the final state M+1. This estimate can be based on the "quasi-stationary approximation", widely used in chemical kinetics [7 - 11].

Due to a high free energy $F_i$ of all intermediate states $1 \le i \le M$, the number of molecules $n_i$ in each intermediate state is very small as compared to their number $n_0 + n_{M+1}$ in the initial and final states. Therefore, the rate of change of $n_i$ is very small as compared to the rate of change of $n_0$ or $n_{M+1}$. Thus, in the "zero" approximation, one can assume that $\frac{dn_i}{dt} = 0$ for intermediate states 1,…,M, and therefore the flow rate is constant along the reaction pathway:

$$-dn_0/dt \equiv k_{0 \to 1}n_0 - k_{1 \to 0}n_1 = k_{1 \to 2}n_1 - k_{2 \to 1}n_2 = \ldots = k_{M \to M+1}n_M - k_{M+1 \to M}n_{M+1} = dn_{M+1}/dt.$$

Denoting the flow rate $(-dn_0/dt)$ as $I$, one has a system of equations

$$\left\{ \begin{array}{c} k_{0 \to 1}n_0 - k_{1 \to 0}n_1 = I \\ \ldots \ldots \\ k_{i-1 \to i}n_{i-1} - k_{i \to i-1}n_i = I \\ \ldots \ldots \\ k_{M \to M+1}n_M - k_{M+1 \to M}n_{M+1} = I, \end{array} \right. \tag{1}$$

or

$$\left\{ \begin{array}{c} n_0 - (k_{1 \to 0}/k_{0 \to 1})n_1 = I \ddagger_1 \\ \ldots \ldots \\ n_{i-1} - (k_{i \to i-1}/k_{i-1 \to i})n_i = I \ddagger_i \\ \ldots \ldots \\ n_M - (k_{M+1 \to M}/k_{M \to M+1})n_{M+1} = I \ddagger_{M+1}, \end{array} \right. \tag{2}$$

where $\ddagger_i \equiv 1/k_{i-1 \to i}$ is the passage ("state $i$-1"-to-"state $i$") time.

Multiplying equation for each $\ddagger_i$ (with $i > 1$) by $(k_{1 \to 0}/k_{0 \to 1}) \cdot \ldots \cdot (k_{i-1 \to i-2}/k_{i-1 \to i-2})$ and summing all these equations, one obtains

$$n_0 - (k_{1 \to 0}/k_{0 \to 1}) \cdot \ldots \cdot (k_{M+1 \to M}/k_{M \to M+1})n_{M+1} = [\ddagger_1 + \ldots + \tau_{M+1} \cdot (k_{1 \to 0}/k_{0 \to 1}) \cdot \ldots \cdot (k_{M \to M-1}/k_{M-1 \to M})] \cdot I$$

Using the well-known [10] ratio $k_{i \to j}/k_{j \to i} = \exp\left[\frac{F_i - F_j}{k_B T}\right]$, which follows from that that the equilibrium populations $n_i^0$ and $n_j^0$ of states $i$ and $j$ must satisfy both to the kinetic equation $n_i^0 k_{i \to j} = n_j^0 k_{j \to i}$ and the thermodynamic relation $n_i^0/n_j^0 = \exp\left[-\frac{F_i - F_j}{k_B T}\right]$ (where $T$ is temperature and $k_B$ the Boltzmann constant), one obtains

$$n_0 - \exp\left[\frac{F_{M+1} - F_0}{k_B T}\right]n_{M+1} = [\ddagger_1 + \ldots + \ddagger_{M+1} \cdot \exp\left[\frac{F_M - F_0}{k_B T}\right]] \cdot I.$$

Thus,

$$I = \frac{n_0 - \exp\left[\frac{F_{M+1} - F_0}{k_B T}\right] \cdot n_{M+1}}{\sum_{j=1}^{M+1} \tau_j \exp\left[\frac{F_{j-1} - F_0}{k_B T}\right]}. \tag{3}$$

If $n_{M+1} << n_0$ and $F_{M+1} < F_0$, then $n_0 >> \exp\left[\frac{F_{M+1} - F_0}{k_B T}\right] \cdot n_{M+1}$, so that

$$I \cong \frac{n_0}{\sum_{j=1}^{M+1} \tau_j \exp\left[\frac{F_{j-1} - F_0}{k_B T}\right]} ; \tag{4}$$

Thus, our task, in fact, is reduced to the calculation of the flux in an irreversible reaction

The characteristic time of passage of all $n_0$ particles from the initial state 0 to M+1 is

$$t_{0 \to \ldots \to M+1} = n_0 / I \cong \sum_{j=1}^{M+1} \tau_j \exp\left[\frac{F_{j-1} - F_0}{k_B T}\right]. \tag{5}$$

Here $F_{j-1} - F_0$ is the free energy state $j$-1 counted off the free energy of the initial state 0. Note that the intermediates of a high free energy make a major contribution to the passage time, and that intermediates of a very low free energy (see the right parts of Figs. 1, 2) make a such a low contribution to the passage time that it can be neglected.

One can also obtain populations $n_i$ of intermediate states in the course of reaction, using the recurrence relations following from the system (2) and equation (4):

$$n_i = (k_{i-1 \to i} / k_{i \to i-1}) \cdot n_{i-1} - (k_{i-1 \to i} / k_{i \to i-1}) \cdot \tau_i \cdot I = \exp\left[\frac{F_{i-1} - F_i}{k_B T}\right] \cdot \{n_{i-1} - \tau_i \cdot I\}$$

$$= \exp\left[\frac{F_0 - F_i}{k_B T}\right] \cdot n_0 - I \cdot \sum_{j=1}^{i} \tau_j \exp\left[\frac{F_{j-1} - F_i}{k_B T}\right]$$

$$= \exp\left[\frac{F_0 - F_i}{k_B T}\right] \cdot \left\{ n_0 - \exp\left[\frac{F_i - F_0}{k_B T}\right] \cdot \frac{n_0}{\sum_{j=1}^{M+1} \tau_j \exp\left[\frac{F_{j-1} - F_0}{k_B T}\right]} \cdot \sum_{j=1}^{i} \tau_j \exp\left[\frac{F_{j-1} - F_i}{k_B T}\right] \right\}$$

$$= \exp\left[\frac{F_0 - F_i}{k_B T}\right] \cdot n_0 \cdot \frac{\sum_{j=i+1}^{M+1} \tau_j \exp\left[\frac{F_{j-1} - F_0}{k_B T}\right]}{\sum_{j=1}^{M+1} \tau_j \exp\left[\frac{F_{j-1} - F_0}{k_B T}\right]} \tag{6}$$

for all $i = 0, 1, \ldots, M+1$. Note that any $n_i$ may be represented as

$$n_i = n_i^0 \cdot \frac{\sum_{j=i+1}^{M+1} \tau_j \exp\left[\frac{F_{j-1} - F_0}{k_B T}\right]}{\sum_{j=1}^{M+1} \tau_j \exp\left[\frac{F_{j-1} - F_0}{k_B T}\right]} , \tag{7}$$

where $n_i^0 = \exp\left[\frac{F_0 - F_i}{k_B T}\right] \cdot n_0$ is population of state $i$, corresponding to its thermodynamic equilibrium with population $n_0$ of the initial state 0.

One can see that all $n_i < n_i^0$ at $i > 0$.

However, the solution obtained above seems to be not quite correct, since the system (2) is obtained under the assumption that $\frac{dn_i}{dt} = 0$ for all intermediates $i = 1, \ldots, M$, and that only $\frac{dn_0}{dt} = -\frac{dn_{M+1}}{dt} = -I \neq 0$. On the other hand, equation (6) leads to non-zero derivatives $\frac{dn_i}{dt}$ for all intermediate states $i = 1, \ldots, M$:

$$\frac{dn_i}{dt} = \exp\left[\frac{F_0 - F_i}{k_B T}\right] \cdot \frac{dn_0}{dt} \cdot \frac{\sum_{j=i+1}^{M+1} \tau_j \exp\left[\frac{F_{j-1} - F_0}{k_B T}\right]}{\sum_{j=1}^{M+1} \tau_j \exp\left[\frac{F_{j-1} - F_0}{k_B T}\right]} = -I \exp\left[\frac{F_0 - F_i}{k_B T}\right] \cdot \frac{\sum_{j=i+1}^{M+1} \tau_j \exp\left[\frac{F_{j-1} - F_0}{k_B T}\right]}{\sum_{j=1}^{M+1} \tau_j \exp\left[\frac{F_{j-1} - F_0}{k_B T}\right]} = -I \cdot \frac{n_i}{n_0}. \tag{8}$$

It is therefore necessary to consider the next, higher approximation taking into account the possible change of population of the intermediate states $i$ (and, consequently, the possible change in the magnitude of the flux $I_i$) at different steps of the above shown irreversible reaction:

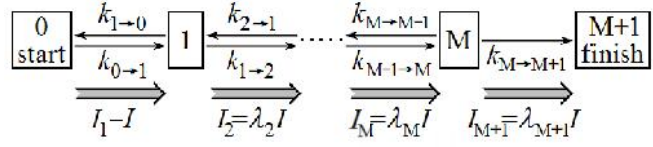

Thus, for all $i = 1, \dots, M+1$ one obtains equations analogous to those presented in system (1),

$$k_{i-1 \to i} n_{i-1} - k_{i \to i-1} n_i = \}_i I \qquad \text{(where } I \equiv -\frac{dn_0}{dt} \text{ and the multiplier } \}_1 \equiv 1), \tag{1a}$$

and the recursive relations similar to those presented in system (2) and equation (6):

$$n_i = \exp\left[\frac{F_{i-1} - F_i}{k_B T}\right] \cdot [n_{i-1} - (\ddagger_i \cdot \}_i) I]. \tag{2a}$$

After calculations similar to those done to derive equations (4), (6), one obtains:

$$I \cong \frac{n_0}{\sum_{j=1}^{M+1} \ddagger_j \lambda_j \exp\left[\frac{F_{j-1} - F_0}{k_B T}\right]} \tag{4a}$$

and

$$n_i = \exp\left[\frac{F_0 - F_i}{k_B T}\right] \left\{ n_0 1 \frac{\sum_{j=i+1}^{M+1} \tau_j \lambda_j \exp\left[\frac{F_{j-1} - F_0}{k_B T}\right]}{\sum_{j=1}^{M+1} \ddagger_j \lambda_j \exp\left[\frac{F_{j-1} - F_0}{k_B T}\right]} \right\} \tag{6a}$$

or

$$n_i = n_i^0 1 \frac{\sum_{j=i+1}^{M+1} \tau_j \lambda_j \exp\left[\frac{F_{j-1} - F_0}{k_B T}\right]}{\sum_{j=1}^{M+1} \ddagger_j \lambda_j \exp\left[\frac{F_{j-1} - F_0}{k_B T}\right]}. \tag{7a}$$

(cf. (6), (7)). Up to now, we did not do any approximations. Now we introduce a stationary approximation, i.e., we assume that each $\frac{d\lambda_i}{dt} = 0$ (without this approximation, we would have to solve the complete system of linear differential equations considered in [6]). As a result (cf. (8)),

$$\frac{dn_i}{dt} = \exp\left[\frac{F_0 - F_i}{k_B T}\right] 1 \frac{dn_0}{dt} 1 \frac{\sum_{j=i+1}^{M+1} \tau_j \lambda_j \exp\left[\frac{F_{j-1} - F_0}{k_B T}\right]}{\sum_{j=1}^{M+1} \ddagger_j \lambda_j \exp\left[\frac{F_{j-1} - F_0}{k_B T}\right]} = -I \cdot \frac{n_i}{n_0}, \tag{8a}$$

where $I \equiv -\frac{dn_0}{dt}$ is now described by equation (4a).

Equation (1a) and the above given scheme show that the flux

$$\}_i I = -\frac{d}{dt}\left(\sum_{j=0}^{i-1} n_j\right). \tag{9}$$

This means that $\}_i I = I \cdot \sum_{j=0}^{i-1} \frac{n_j}{n_0}$, or

$$\}_i = 1 + \sum_{j=1}^{i-1} \frac{n_j}{n_0}, \tag{10}$$

i.e., $\}_i$ increases with $i$.

However, since the population of each state, $n_j$, in the course of reaction does not exceed $n_j^0$, the thermodynamically equilibrium population of the same state (see equation (7a)),

$$\}_i \le 1 + \sum_{j=1}^{i-1} \frac{n_j^0}{n_0} \equiv 1 + \sum_{j=1}^{i-1} \exp\left[\frac{F_0 - F_j}{k_B T}\right], \tag{11}$$

which means that all the values $\}_i$ remain close to 1 if $\sum_{j=1}^{i-1} \exp\left[\frac{F_0 - F_i}{k_B T}\right] << 1$, i.e., if the free energies of the intermediate states $j = 1, \ldots, i\text{-}1$ are much, by many $k_B T$ higher than $F_0$.

This means that the estimate of characteristic reaction time obtained in equation (5) (and in (8.19) of [5]) is fairly accurate, *provided* that the free energy barrier at the reaction pathway is high.

In conclusion, I have to repeat that I was surprised when failed to find in literature a general formula (5) to estimate the characteristic time of crossing of arbitrarily bumpy but high free energy barrier, and I will be obliged to any reader who will send me the corresponding reference(s).

**Acknowledgements**

I am grateful to Adela Croce and Gert Van der Zwan who kindly helped me to find some important references. The work has been supported in part by RFBR (13-04-00253a), MCB RAS (01201358029) and MES RK grants.